\documentclass[aps,prb,twocolumn,superscriptaddress]{revtex4-1}
\usepackage{dcolumn,amsfonts,amsmath,amssymb}
\usepackage{bm}

\usepackage{color}

\usepackage{graphicx}

\newcommand{\kup}{|\!\uparrow\rangle}
\newcommand{\kdn}{|\!\downarrow\rangle}
\newcommand{\ktup}{|T\!\uparrow\rangle}
\newcommand{\ktdn}{|T\!\downarrow\rangle}
\newcommand{\bup}{\langle\uparrow\!|}
\newcommand{\bdn}{\langle\downarrow\!|}
\newcommand{\btup}{\langle T\!\uparrow\!|}

\newcommand{\wt}{\omega_{\mathrm{t}}}
\newcommand{\wh}{\omega_{\mathrm{h}}}
\newcommand{\St}{\Sigma_{\mathrm{t}}}
\newcommand{\Sh}{\Sigma_{\mathrm{h}}}
\newcommand{\gt}{\gamma_{\mathrm{t}}}
\newcommand{\gh}{\gamma_{\mathrm{h}}}
\newcommand{\gR}{\gamma_{\mathrm{R}}}

\begin{document}
\title{Decoherence-assisted initialization of a resident hole spin
  polarization in a p-doped semiconductor quantum well}
\author{M.\ Kugler}
\affiliation{Institut f\"ur Experimentelle und Angewandte Physik,
Universit\"at Regensburg, D-93040 Regensburg, Germany}
\author{K. Korzekwa}
\affiliation{Institute of Physics, Wroc{\l}aw University of
  Technology, 50-370 Wroc{\l}aw, Poland}
\author{P. Machnikowski}\email{Pawel.Machnikowski@pwr.wroc.pl}
\affiliation{Institute of Physics, Wroc{\l}aw University of
  Technology, 50-370 Wroc{\l}aw, Poland}
\author{C.\ Gradl}
\affiliation{Institut f\"ur Experimentelle und Angewandte Physik,
Universit\"at Regensburg, D-93040 Regensburg, Germany}
\author{S.\ Furthmeier}
\affiliation{Institut f\"ur Experimentelle und Angewandte Physik,
Universit\"at Regensburg, D-93040 Regensburg, Germany}
\author{M.\ Griesbeck}
\affiliation{Institut f\"ur Experimentelle und Angewandte Physik,
Universit\"at Regensburg, D-93040 Regensburg, Germany}
\author{M.\ Hirmer}
\affiliation{Institut f\"ur Experimentelle und Angewandte Physik,
Universit\"at Regensburg, D-93040 Regensburg, Germany}
\author{D.\ Schuh}
\affiliation{Institut f\"ur Experimentelle und Angewandte Physik,
Universit\"at Regensburg, D-93040 Regensburg, Germany}
\author{W.\ Wegscheider}
\affiliation{Solid State Physics Laboratory, ETH Zurich, 8093 Zurich, Switzerland}
\author{T.\ Kuhn}
\affiliation{Institut f\"ur Festk\"orpertheorie, Westf\"alische Wilhelms-Universit\"at, D-48149 M\"unster, Germany}
\author{C.\ Sch\"uller}
\affiliation{Institut f\"ur Experimentelle und Angewandte Physik,
Universit\"at Regensburg, D-93040 Regensburg, Germany}
\author{T.\ Korn}
\email{tobias.korn@physik.uni-r.de} \affiliation{Institut
f\"ur Experimentelle und Angewandte Physik, Universit\"at
Regensburg, D-93040 Regensburg, Germany}
\date{\today}

\begin{abstract}
We  investigate  spin dynamics of resident holes in
a p-modulation-doped GaAs/Al$_{0.3}$Ga$_{0.7}$As single quantum well.
Time-resolved Faraday and Kerr rotation, as well as resonant spin
amplification, are utilized in our study.
We observe that nonresonant or high power optical pumping leads to a resident hole
spin polarization with opposite sign with respect to the optically
oriented carriers, while low power resonant
optical pumping only leads to a resident hole spin polarization if a
sufficient in-plane magnetic field is applied. The competition between
two
different processes of spin orientation strongly modifies the shape of resonant spin
amplification traces. Calculations of the spin dynamics in the
electron--hole system are in good agreement with the experimental Kerr
rotation and resonant spin amplification traces and allow us to
determine the hole spin polarization within the sample after optical
orientation, as well as to extract quantitative information about
spin dephasing processes at various stages of the evolution.
\end{abstract}

\maketitle

\section{Introduction}
The promising field of semiconductor spintronics~\cite{fabian07} has
stimulated a large number of studies of spin dynamics and spin
dephasing mechanisms in a vast variety of semiconductors and their
heterostructures in recent years.  Especially the dynamics of
conduction-band electrons in compound semiconductors without inversion
center, e.g., GaAs, have been studied by many groups (see, e.g.,
Ref.~\onlinecite{WuReview} for a recent review). By contrast, hole
spin
dynamics in these systems have been investigated with less
intensity. This is, in part, due to the sub-picosecond hole spin
dephasing time (SDT) in bulk GaAs~\cite{hilton02,wu2010}, which arises from
the strong spin-orbit coupling within the p-like valence bands.  In
p-doped quantum wells (QWs), an \emph{increase} of the hole SDT to a
few picoseconds has  been observed experimentally by several
groups~\cite{damen91,ganichev04} and reproduced in microscopic
calculations~\cite{PhysRevB.73.125314}. Significantly longer hole SDT
have been observed for optically oriented holes in n-doped
QWs~\cite{marie95,marie99}, and more recently in p-doped QW systems in
which localization of holes occurs at low
temperatures~\cite{syperek07, Korn10,KornReview,Studer11}, and also in quantum
dots~\cite{heiss07}. For localized electrons in quantum dots, the
contact hyperfine interaction leads to ensemble spin dephasing on the
10~ns scale~\cite{petta05}. However, due to their p-like wave functions,
this dephasing process is suppressed for holes, and only the weaker
dipole-dipole interaction has to be taken into
account~\cite{fischer:155329}. Therefore, localized holes may be more
suitable than electrons for the realization of future quantum
computing schemes. Additionally, the large orientational anisotropy of
the hole $g$ factor in GaAs-based heterostructures~\cite{winkler00}
strongly influences hole spin dynamics in tilted magnetic
fields~\cite{Kuhn10} and may allow for spin manipulation schemes based
on electrical $g$ factor
control~\cite{PhysRevB.79.045307,kugler:035325}.

Here, we present time-resolved studies of the combined electron and
hole spin dynamics in a p-modulation doped quantum well
under different excitation conditions. We utilize
time-resolved Kerr/Faraday rotation~\cite{Baumberg_Faraday},  as well as the
related resonant spin amplification technique~\cite{awschalom98}, and
time-resolved photoluminescence. We 
identify two processes in which spin polarization is transferred to
the resident holes after optical orientation,  quantitatively model
the dynamics, and determine the contributions of these processes
depending on excitation conditions. By fitting  our theoretical model to the experimental
results, we are able to find the degree of spin
polarization after the optical excitation and to extract the hole spin
dephasing times, as well as the degree of coherence loss during
excitation at various conditions of optical pumping. We determine transverse spin dephasing times $T_2$ of almost 100~ns under weak, resonant excitation. Additionally, we
show that fast dephasing of the hole spin state during and just after
high-power or blue-detuned pumping leads to polarization of the hole
spins at zero magnetic field, while at finite fields this process
competes with the polarization mechanism due to trion spin precession
\cite{syperek07,kugler:035325}. As these two competing processes lead to
opposite spin orientation our findings show that the spin orientation
can be controlled by modifying the optical excitation conditions.

The paper is organized as follows. First, in Sec.~\ref{sec:sample}, we
present the sample and the idea of the experiment. Next, in
Sec.~\ref{sec:theor}, the theoretical model is introduced. Sec.~\ref{sec:results} contains the presentation and
discussion of the experimental results and their theoretical
modeling. Finally, Sec.~\ref{sec:concl} concludes the paper.

\section{Sample structure and experimental methods}
\label{sec:sample}

Our samples are single-side p-modulation-doped
GaAs/Al$_{0.3}$Ga$_{0.7}$As QWs (QW width 4~nm), containing a
two-dimensional hole system (2DHS) 
with a hole density $p = 1.1 \times 10^{11}$~cm$^{-2}$ and mobility
$\mu = 1.3 \times 10^{4}$~cm$^2/$Vs (measured at 1.3~K) from a single
wafer grown by 
molecular beam epitaxy. Some samples from this wafer are thinned for
measurements in transmission. For this, the samples are first
glued onto a sapphire substrate with optically transparent glue, then
the semiconductor substrate is removed by grinding and selective wet
etching. The samples contain a short-period GaAs/AlGaAs superlattice,
which serves as an etch stop, leaving only the MBE-grown
layers. Earlier studies of spin dynamics performed on similar systems
\cite{syperek07,kugler:035325} indicate that in structures of this kind
the resident holes are weakly trapped, most likely on QW width
fluctuations. This is confirmed by a rapid increase of hole spin
dephasing above a certain threshold temperature, associated with
thermal release of the carriers from these binding centers and the
onset of spin-orbit-related dephasing characteristic of free carriers.

The   resonant spin amplification (RSA) measurements are performed in
an optical cryostat with $^3$He insert, allowing us to lower the sample
temperatures below 400~mK and to apply magnetic fields of up to
11.5~Tesla. Here, the samples are cooled by cold $^3$He gas. For some of these
measurements, thinned samples are used and the experiment is
performed in transmission (Faraday rotation) to limit the amount of
absorbed laser power. The time-resolved Kerr rotation (TRKR)
measurements are performed in a Helium flow cryostat, in which the
samples are mounted on the cold finger of the cryostat in vacuum.  A
pulsed Ti-Sapphire laser system generating pulses with a length of
600~fs and a spectral width of 3-4~meV is used for the optical
measurements. The repetition rate of the laser system is 80~MHz,
corresponding to a time delay of 12.5~ns between subsequent
pulses. The laser pulses are split into a circularly-polarized pump
beam and a linearly-polarized probe beam by a beam splitter. A
mechanical delay line is used to create a variable time delay between
pump and probe. Both beams are focused to a diameter of about
80~$\mu$m on the sample using an achromat.

In the TRKR and RSA experiments, the circularly-polarized pump beam is
generating electron-hole pairs in the QW, with spins aligned parallel
or antiparallel to the beam direction, i.e., the QW normal,
depending on the helicity of the light.  In the
TRKR measurements, the spin polarization created perpendicular to the
sample plane by the pump beam, is probed by the time-delayed probe
beam via the Kerr effect: the axis of linear polarization of the probe
beam is rotated by a small angle, which is proportional to the
out-of-plane component of the spin polarization
\cite{Kuhn10,yugova09}.   This small angle is
detected using an optical bridge. A lock-in scheme is used to increase
sensitivity. The RSA technique is based on the
  interference of spin polarizations created in a sample by subsequent
  pump pulses. It requires that the spin dephasing time is comparable
  to the time delay between pump pulses. For certain magnetic fields
  applied in the sample plane, the optically oriented spin
  polarization precesses by an integer multiple of
  $2\pi$ in the time
  window between subsequent pump pulses, so that constructive
  interference occurs. This leads to pronounced maxima in the Faraday
  or Kerr rotation angle  measured
for a fixed time delay as a function of the applied  magnetic
field.  In our measurements, the time delay is chosen to probe the spin polarization
remaining within the sample 100~ps before the arrival of a pump pulse.

Time-resolved photoluminescence (TRPL) measurements
  are performed using a Hamamatsu streak camera system synchronized to
  the pulsed Ti-Sapphire laser system. For these measurements, the
  laser is detuned to create electron-hole pairs at an energy about
  30~meV above the photoluminescence (PL) energy of the heavy-hole
  exciton and trion lines. The PL from the sample is collected using
  an achromat and dispersed in a spectrometer before being detected by
  the streak camera.

\begin{figure}
\centering
\includegraphics[width=85mm]{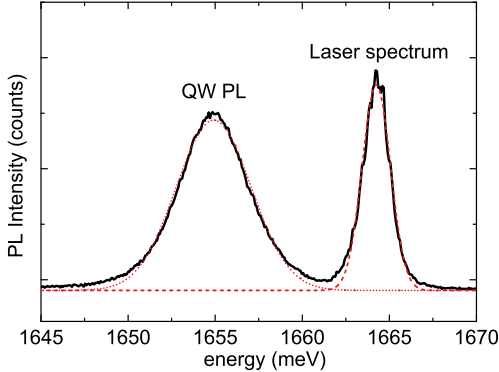}
\caption{\label{Fig0_PL}(Color online)
Photoluminescence (PL) trace of the sample measured at 1.2~K and
spectrum of the  Ti-Sapphire laser system used in the time-resolved
experiments. The broken lines show Gaussian fits to the quantum well
PL (dotted line) and the laser spectrum (dashed line).} 
\end{figure}

For initial characterization of the samples, PL measurements using
continuous-wave excitation with a 532~nm laser are performed. A
grating spectrometer with a Peltier-cooled charge coupled device (CCD)
detector is used to collect the PL. Figure~\ref{Fig0_PL} shows a
typical PL trace measured at a sample temperature of 1.2~K. The PL
from the QW is a near-symmetrical single peak with a spectral width of
about 5~meV. No fine structure of this peak, corresponding to, e.g.,
neutral or positively charged excitons, is observed in PL
measurements. In the same figure, a typical spectrum of the
Ti-Sapphire laser system can be seen at higher energy. 
In the following, by ``resonant excitation'' we will understand the
situation where the Ti-sapphire laser system is tuned to achieve
maximum Kerr signal. Since our sample is doped,  the states
corresponding to the maximum of the PL emission are occupied by
resident holes, therefore, the absorption at this energy is
suppressed. Maximum Kerr signal is observed for an energetic position
of the laser at the high-energy flank of the QW PL emission. 

\section{Theoretical model}
\label{sec:theor}

In order to interpret the experimental results
we propose a minimal, generic model that is able to account for all
the features of the spin dynamics observed in the experiment without
specific assumptions on the detailed mechanism of spin decoherence.
In accordance with the previous experimental
findings\cite{syperek07,kugler:035325}, we
assume that the optical response can be described in terms of
independent hole-trion systems, trapped in QW fluctuations.
The
state of each such system is represented by the density matrix $\rho$,
restricted to the four relevant states \mbox{$\kup,\kdn,\ktup,\ktdn$},
representing the two hole states and the two trion states with
different spin orientations (with respect to the system symmetry axis,
normal to the sample plane). We neglect the influence of the weak probe
pulse on the spin dynamics and calculate the spin polarization at the
arrival of the probe, which is known\cite{Kuhn10,yugova09} to be
translated by the probe into the Kerr or Faraday signal.

The experiment is modeled by a sequence of three steps:

First, the pump pulse transforms the initial state $\rho_{0}$ into a
new state $\rho_{1}$, described up to the second order in the pulse
amplitude by
\begin{eqnarray}
\rho_{1} & = &
-\frac{i}{\hbar}\int_{-\infty}^{\infty} dt
\left[H_{\mathrm{l}}(t),\rho_{0}\right] \nonumber \\
&&-\frac{1}{\hbar^{2}}\int_{-\infty}^{\infty}dt\int_{-\infty}^{\infty}dt'
\left[H_{\mathrm{l}}(t),
  \left[H_{\mathrm{l}}(t'),\rho_{0}\right]\right],
\label{laser}
\end{eqnarray}
where we use the fact that the pulse is very short compared to the
spin evolution time scales and assume, for simplicity, that the
excitation is coherent. Here,
$H_{\mathrm{l}}=(1/2)f(t)\mbox{$\kup\!\btup$}+\mathrm{h.c.}$ is the
carrier-laser coupling Hamiltonian with a pulse envelope
$f(t)$ and $\sigma^+$ circular polarization is assumed for the laser pulse.

Second, we allow for a fast partial decoherence of the hole spin which takes
place on time scales much shorter than the subsequent spin dynamics
and is therefore modeled as instantaneous. This leads to a system
state $\rho_{2}$ with
\begin{displaymath}
\bdn\rho_{2}\kup
=\bdn\rho_{1}\kup e^{-u/2-w}
\end{displaymath}
and
\begin{displaymath}
\left\langle \alpha|\rho_{2}|\alpha\right\rangle=
\frac{1}{2}\mbox{$\bup\rho_{1}\kup$}\left( 1\pm e^{-u} \right)
+\frac{1}{2}\bdn\rho_{1}\kdn\left( 1\mp e^{-u} \right),
\end{displaymath}
where $\alpha=\uparrow,\downarrow$ and the upper (lower) sign is for
$\alpha=\uparrow\!(\downarrow)$. The factors $e^{-u}$ and $e^{-w}$
describe the effects of occupation relaxation and additional pure
dephasing in the reference frame associated with the system symmetry
axis (coinciding with the axis of optical orientation), which is the
relevant one here in view of the fast character of the
process, as compared to the Larmor precession.

It should be noted that the combination of coherent excitation and
instantaneous dephasing does not necessarily reflect the actual
microscopic kinetics of the system. In particular, for off-resonant
excitation, hole spin flips during relaxation to low-energy states are
also possible. While this process is clearly beyond our four-level
model, its essential effect is bringing the hole spin polarization
towards equilibrium and dephasing of the hole spin coherence. Both
these effects are included in our model in terms of the instantaneous
relaxation and dephasing factors $e^{-u}$ and $e^{-w}$.

In the third stage, the system evolution (Larmor precession,
recombination and spin
decoherence) is modeled in terms of the Markovian Master equation
(in the Schr\"odinger picture with respect to the spin dynamics but in the
rotating frame with respect to the interband transition energy)
\begin{equation}\label{evol}
\dot{\rho}=-\frac{i}{\hbar}[H_{0},\rho]
+\mathcal{L}_{\mathrm{h}}[\rho]+\mathcal{L}_{\mathrm{t}}[\rho]
+\mathcal{L}_{\mathrm{r}}[\rho],
\end{equation}
with the initial condition $\rho(0)=\rho_{2}$.
Here
\begin{equation*}
H_{0}=-\frac{1}{2}\mu_{\mathrm{B}}\bm{B}\hat{g}_{\mathrm{h}}\bm{\sigma}_{\mathrm{h}}
-\frac{1}{2}g_{\mathrm{t}}\mu_{\mathrm{B}}\bm{B}\cdot\bm{\sigma}_{\mathrm{t}},
\end{equation*}
is the hole and trion spin Hamiltonian,
where $\mu_{\mathrm{B}}$ is the Bohr magneton,
$\hat{g}_{\mathrm{h}}$ is the hole Land{\'e} tensor,
$g_{\mathrm{t}}$ is the Land{\'e} factor of the trion
(i.e., essentially, of the electron), which we assume to be isotropic, and
$\bm{\sigma}_{\mathrm{h}},\bm{\sigma}_{\mathrm{t}}$ are the vectors of
Pauli matrices corresponding to the hole and trion spin, respectively
(the hole is treated as a pseudo-spin-1/2 system),
in the basis of hole spin states $\mbox{$\kup,\kdn $}$ and trion spin
states $\mbox{$\ktup,\ktdn $}$. This Hamiltonian accounts for the spin
pression with the Larmor frequencies 
$\wh=\mu_{\mathrm{B}}|\hat{g}_{\mathrm{h}}\bm{B}|/\hbar$ and 
$\wt=\mu_{\mathrm{B}}g_{\mathrm{t}}B/\hbar$ for the hole and trion, respectively.

The hole dissipator
$\mathcal{L}_{\mathrm{h}}$ is obtained within the standard
weak-coupling approach \cite{breuer02}
from the hole spin-environment Hamiltonian
$H_{\mathrm{he}}=\bm{\sigma}\cdot\bm{R}^{(\mathrm{h})}$,
where $\bm{R}^{(\mathrm{h})}$ are certain
environment operators. We derive the evolution equation for the
hole spin in the Markov limit as in Ref.~\onlinecite{Kuhn10}
but without the secular approximation which does not hold in the
general case of possibly low or vanishing magnetic fields. As a result, we get
the dissipator (in the Schr\"odinger picture) in the form
\begin{eqnarray*}
\mathcal{L}_{\mathrm{h}}[\rho] & = & -\pi\sum_{lj} \left[
R_{lj}^{(\mathrm{h})} (\omega_{j})\left( \sigma_{l}\sigma_{j}\rho
-\sigma_{j}\rho\sigma_{l} \right)\right.\\
&&\left.+R_{lj}^{(\mathrm{h})} (-\omega_{l})\left( \rho  \sigma_{l}\sigma_{j}
-\sigma_{j}\rho\sigma_{l} \right)  \right],
\end{eqnarray*}
where $l,j=\pm,0$, $\omega_{0}=0$,
$\omega_{+}=-\omega_{-}=\omega_{\mathrm{h}}$, and
$\sigma_{\pm,0}$ are Pauli matrices in the reference frame associated
with the $x$ direction (the orientation of the field),
\begin{displaymath}
\sigma_{0}=\sigma_{x},\quad
\sigma_{+}=\sigma_{-}^{\dag}=\frac{-\sigma_{z}+i\sigma_{y}}{2}.
\end{displaymath}
The spectral densities for the hole reservoir are
\begin{displaymath}
R_{lj}^{(\mathrm{h})} (\omega)=\frac{1}{2\pi\hbar^{2}}\int dt e^{i\omega t}
\langle R_{l}^{(\mathrm{h})} (t)R_{j}^{(\mathrm{h})}\rangle,
\quad l,j=\pm,0,
\end{displaymath}
where
$R_{0}^{(\mathrm{h})}=R_{x}^{(\mathrm{h})}$,
$R_{+}^{(\mathrm{h})}=R_{-}^{(\mathrm{h})\dag}=-R_{z}^{(\mathrm{h})}-iR_{y}^{(\mathrm{h})}$,
and $R_{l}^{(\mathrm{h})} (t)$ denotes the operator in the interaction picture
with respect to the environment Hamiltonian.
Consistently with the assumed $C_{4\mathrm{v}}$ symmetry of the
system, we  set
$R_{\alpha\beta}^{(\mathrm{h})}
(\omega)=0$ for $\alpha,\beta=x,y,z$,
$\alpha\neq\beta$ and
$\quad R_{yy}(\omega)=R_{xx}(\omega)$.
The trion spin dissipator
$\mathcal{L}_{\mathrm{t}}$ is obtained in the same way
with a set of trion-related spectral densities
$R_{\alpha\beta}^{(\mathrm{t})}(\omega)$. We assume that
the reservoirs coupled to electron (trion) and hole spins are
uncorrelated.

The last term in Eq.~\eqref{evol} is the standard spontaneous emission
generator that accounts for the radiative
recombination of the trion (see Ref.~\onlinecite{Kuhn10}) with the
rate $\gR$.

Eq.~\eqref{evol} can be rewritten in terms of the three components of
the hole spin polarization
\begin{align*}
X_{\mathrm{h}}&=\bup\rho\kdn+\bdn\rho\kup,&
Y_{\mathrm{h}}&=i(\bup\rho\kdn-\bdn\rho\kup),\\
\Sigma_{\mathrm{h}}&=\bup\rho\kup-\bdn\rho\kdn
\end{align*}
(and analogous for the trion). For the hole spin polarization, the
equations of motion are
\begin{subequations}
\begin{eqnarray}
\dot{X}_{\mathrm{h}}& = &
-\left( \kappa_{z} + \kappa_{x} \right) X_{\mathrm{h}}
+(\kappa'_{x}+\kappa'_{z}) N_{\mathrm{h}}, \label{evol-X}\\
\dot{Y}_{\mathrm{h}} & = & \wh \Sigma_{\mathrm{h}}
-\left( \kappa_{x0} +\kappa_{z} \right) Y_{\mathrm{h}}, \label{evol-Y}\\
\dot{\Sigma}_{\mathrm{h}} & = & -\wh Y_{\mathrm{h}}
 -\left( \kappa_{x} + \kappa_{x0} \right)\Sigma _{\mathrm{h}}
+\gR \Sigma_{\mathrm{t}}, \label{evol-Z}
\end{eqnarray}
\end{subequations}
where $N_{\mathrm{h}}$ is the hole population and
\begin{subequations}
\begin{align}
\kappa_{\alpha} & = 2\pi \left[
R_{\alpha\alpha}^{(\mathrm{h})}(\omega_{\mathrm{h}})
+R_{\alpha\alpha}^{(\mathrm{h})}(-\omega_{\mathrm{h}}) \right], &
\kappa_{\alpha 0} & = 4\pi R_{\alpha\alpha}^{(\mathrm{h})}(0), \label{kappa1}\\
\kappa'_{\alpha} & = 2\pi \left[
R_{\alpha\alpha}^{(\mathrm{h})} (\omega_{\mathrm{h}})
-R_{\alpha\alpha}^{(\mathrm{h})} (-\omega_{\mathrm{h}}) \right], \label{kappa2}
\end{align}
\end{subequations}
for $\alpha=x,z$.
In order to find an interpretation of the dephasing rates appearing in
Eqs.~\eqref{evol-X}--\eqref{evol-Z} we note that at $\bm{B}=0$ one has
$\wh=0$, hence
$\kappa_{\alpha}=\kappa_{\alpha 0}$ and the
decoherence time for
the spin polarization along the structure axis is
$T_{z}^{(0)}=1/(2\kappa_{x 0})$, while the decoherence time for the
in-plane components of the spin polarization is
$T_{xy}^{(0)}=1/(\kappa_{z}+\kappa_{x0})$. On the other hand, in
sufficiently strong fields
(for $\wh\gg \kappa_{\alpha},\kappa_{\alpha  0}$),  the longitudinal
(with respect to the field orientation) spin relaxation time is
$T_{1}=1/(\kappa_{z}+\kappa_{x})$ and the transverse relaxation
(dephasing) time is $T_{2}=2/(\kappa_{z}+\kappa_{x}+2 \kappa_{x0})$.

The equations of motion for the trion spin polarization are
\begin{subequations}
\begin{eqnarray}
\dot{X}_{\mathrm{t}}& = &
-\left( \mu_{z} + \mu_{x} \right) X_{\mathrm{t}}
+(\mu'_{x}+\mu'_{z}) N_{\mathrm{t}}-\gR X_{\mathrm{t}}, \label{evol-t-X}\\
\dot{Y}_{\mathrm{t}} & = & \wt \Sigma_{\mathrm{t}}
-\left( \mu_{x0} +\mu_{z} \right) Y_{\mathrm{t}}-\gR Y_{\mathrm{t}}, \label{evol-t-Y}\\
\dot{\Sigma}_{\mathrm{t}} & = & -\wt Y_{\mathrm{t}}
 -\left( \mu_{x} + \mu_{x0} \right)\Sigma _{\mathrm{t}}
-\gR \Sigma_{\mathrm{t}}, \label{evol-t-Z}
\end{eqnarray}
\end{subequations}
where $\mu_{\alpha}$, $\mu_{\alpha}'$, and $\mu_{\alpha 0}$ are the
trion decoherence rates defined as in Eqs.~\eqref{kappa1} and
\eqref{kappa2}, but with the trion-related spectral densities
$R_{\alpha\alpha}^{(\mathrm{t})}(\omega)$ taken at the trion Larmor
frequency $\wt$, and
$N_{\mathrm{t}}$ is the trion occupation.

The optical response, that is, the rotation of the polarization plane
of the reflected or transmitted probe pulse, is proportional
to\cite{yugova09,Kuhn10}
\begin{displaymath}
\Delta\Sigma=\Sigma_{\mathrm{t}}-\Sigma_{\mathrm{h}}.
\end{displaymath}

As the time-resolved Kerr response is investigated for
  experimental conditions of relatively high spin
dephasing and relaxation rates one can assume that the evolution after
each laser
repetition is independent and starts from the thermal equilibrium
state. After the pump pulse, the trion and hole spin polarizations
are
\begin{displaymath}
\Sigma_{\mathrm{t}}=-\Sigma_{\mathrm{h}}=\Sigma^{(0)}.
\end{displaymath}
 As a consequence of the initial dephasing, the hole spin polarization is
reduced to
\begin{equation}\label{fast-relax}
\Sigma_{\mathrm{h}}^{(0)}=-\Sigma^{(0)}e^{-u}
\end{equation}
(we assume no fast dephasing of the trion spin
polarization). Then, by
solving Eq.~\eqref{evol} one gets the Kerr signal at $\bm{B}=0$
in the form
\begin{equation}
\Delta\Sigma^{(\mathrm{Kerr})}=ae^{-\gt t}-be^{-\gh t},
\label{kerr}
\end{equation}
 where
$a=(1+ \eta)\St^{(0)}$,  $b=\Sh^{(0)}+\eta\St^{(0)}$,
$\eta=\gR/[\gt-\gh]$.
Here
$\gamma_{\mathrm{h}}=\kappa_{x}+\kappa_{x0}$ is the hole spin
decoherence rate and
$\gamma_{\mathrm{t}}=\mu_{x}+\mu_{x0}+\gR$ is the trion spin
decoherence rate. Since we do not propose any specific microscopic
mechanism for
the spin decoherence the rates $\gamma_{\mathrm{t}}$,
$\gamma_{\mathrm{h}}$, and $\gamma_{\mathrm{R}}$ are treated as
independent parameters of the model.

For the RSA signal, the spin polarization surviving between subsequent laser
repetitions is essential. In order to find the resonantly amplified
spin polarization just before the pump pulse, we find the
mapping of the hole spin-related variables
$X_{\mathrm{h}},Y_{\mathrm{h}},\Sigma_{\mathrm{h}}$
corresponding to the three-step
state transformation described above, assuming that
trion occupations and interband coherences decay completely in the
repetition interval. Moreover, the RSA measurements are performed
under conditions of long spin dephasing time, hence we assume that the
hole spin dephasing rates are small compared to the trion recombination
rate.
The RSA signal is then found as the fixed point of this three-step transformation
to the leading (second) order in the pulse area, that is, in the weak excitation limit.
The resulting spin polarization just before the arrival of the pump
pulse is proportional to
\begin{subequations}
 \begin{equation}
\Delta\Sigma^{(\mathrm{RSA})}\sim f\frac{P}{Q},
\label{RSA}
\end{equation}
where
\begin{align}
f=&1-e^{-u}-\frac{\wt^{2}}{\gR^{2}+\wt^{2}},\label{RSA1}\\
P=&(i\tilde{\omega}+\kappa')e^{i\tilde{\omega}t_{\mathrm{r}}/2}
-i\tilde{\omega}e^{-u/2-w-\kappa t_{\mathrm{r}}/2}
-(\tilde{\omega}\to -\tilde{\omega}),\label{RSA2}\\
Q=&e^{-u}P+\big[
(i\tilde{\omega}-\kappa')e^{-u/2-w+i\tilde{\omega}t_{\mathrm{r}}/2}
-i\tilde{\omega}e^{\kappa t_{\mathrm{r}}/2} \nonumber \label{RSA3}\\
&-(\tilde{\omega}\to -\tilde{\omega}) \big].
\end{align}
\end{subequations}
 Here, $t_{\mathrm{r}}$ is the laser repetition period,
$\kappa= \kappa_{x}+\kappa_{z}+2\kappa_{x0}$,
$\kappa'=\kappa_{z}-\kappa_{x}$,
$\tilde{\omega}=2\sqrt{\wh^{2}-{\kappa'}^{2}/4}$, and
$(\tilde{\omega}\to -\tilde{\omega})$ represents
additional terms,
obtained from the preceding ones by changing the sign of
$\tilde{\omega}$. It is found
that the RSA response in the weak excitation limit does not directly
depend on detuning.
In order to simulate the response from an inhomogeneous ensemble of
hole spins, the result obtained from Eqs.~\eqref{RSA}--\eqref{RSA3}
was averaged according to a Gaussian distribution of hole $g$-factors
with the standard deviation $\Delta g$.

\section{Results and Discussion}
\label{sec:results}

\begin{figure} \centering
  \includegraphics[width= 0.5\textwidth]{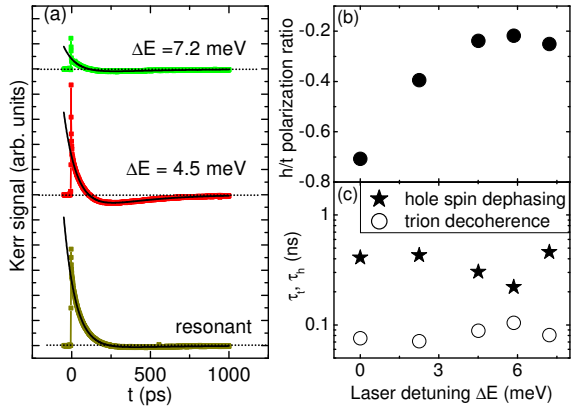}
   \caption{(Color online) (a) TRKR traces measured at 15~K with different laser
excitation energies (symbols). The solid lines represent fits to
the data according to the theory (extended to $t<0$ for better
visibility).  (b) Ratio of hole and trion spin 
polarizations after rapid initial hole dephasing,
$\Sigma_{\mathrm{h}}^{(0)}/\Sigma_{\mathrm{t}}^{(0)}=-e^{-u}$,  extracted from the
measurements as a function of laser detuning. (c) Hole spin dephasing
time (solid stars) and trion spin decoherence time (open symbols) as a
function of laser detuning.}
   \label{TRKR_lambda}
\end{figure}

In this section we present the results of  TRKR and RSA
measurements and interpret them, based on the theoretical model
presented in the previous section. We first discuss the Kerr
measurements at zero magnetic field and then the RSA results.

\subsection{Kerr response at $B=0$}
\label{sec:results-kerr}

First, we investigate the TRKR measurements at zero magnetic
field. Here, we performed two
series of measurements: as a function of the pump frequency detuning and as a
function of the pump pulse power.

In the frequency detuning series,
performed at a fixed sample temperature of 15~K, the
excitation wavelength was tuned from near-resonant excitation to
higher laser energy, which resulted in non-resonant excitation
conditions. Figure \ref{TRKR_lambda}(a) shows three TRKR traces for
resonant excitation and two different values of laser energy
detuning (symbols). While the trace for resonant excitation shows a
near-monoexponential decay, the two traces measured using larger laser
energies display a more complex behavior, with a very rapid initial
decay of the signal and a zero crossing, followed by a slower decay of
the negative signal. Additionally, we note that the Kerr signal
amplitude decreases as the laser energy is detuned from the resonance,
limiting the detuning range accessible in the measurements to about
7~meV. This is due both to a reduced absorption of the pump pulse
and to the spectral dependence of the Kerr rotation of the degenerate
probe pulse.

We interpret the traces as follows: under resonant excitation
conditions in the absence of a magnetic field, both the optically
oriented electrons \emph{and} the
optically oriented holes retain their spin orientation during the
photocarrier lifetime. Electrons recombine
with holes that have a matching spin orientation
according to the selection rules, thereby removing
the optically created spin polarization from the sample during
recombination. Therefore, no hole spin polarization is transferred to
the resident holes. By contrast, under non-resonant excitation
conditions, a fraction of the hole spin polarization is rapidly
lost. This can result from hole spin flips during energy relaxation of
the optically created holes or
from the thermalization of the spin orientation of the resident holes
resulting from binding part of the oppositely oriented holes with the
optically created excitons into trions. On the other hand,
the electrons seem to retain their spin orientation. Upon recombination,
these spin-polarized electrons \emph{remove} holes with matching spin
polarization from the partly depolarized hole system, leaving an excess of
holes oriented opposite to the optically created hole spin
orientation (we will refer to this opposite orientation as
\textit{negative}). We note that both the spin-polarized electrons and
holes
created by interband absorption of circularly polarized light will
lead to the same Kerr rotation of a test beam, so that \emph{a
priori}, the observed Kerr rotation does not allow us to identify the
type of spin-polarized carriers directly. The origin of the Kerr
signal can be determined by applying a magnetic field
perpendicular to the spin polarization and observing the spin
precession, using the different $g$-factors of electrons and holes. Naturally,
in the case of a doped sample like our 2DHS, investigating the Kerr rotation
after photocarrier recombination, so that only the resident carriers
remain, also gives unambiguous results.

The experimental traces are well-reproduced by
Eq.~\eqref{kerr} (the fits are shown as solid lines in
Fig.~\ref{TRKR_lambda}(a)) in the whole time range except for the
first few picoseconds 
after excitation, in which the rapid initial dephasing of the holes
occurs, which is not modeled in a time-resolved manner in the theory. From
the least-squares fit parameters, we are able to extract the ratio of the hole and
electron (trion) spin polarizations,
$\Sigma_{\mathrm{h}}^{(0)}/\Sigma_{\mathrm{t}}^{(0)}=-e^{-u}$ (see Eq.~\eqref{fast-relax})
after the initial dephasing, before
photocarrier recombination occurs. Fig.~\ref{TRKR_lambda}(b) shows
the calculated results as a function of the laser energy detuning from
resonance. We see that, for resonant excitation, the ratio is close to
$-1$, indicating a hole spin polarization almost equal to and oriented
opposite to the electron spin polarization. As the laser energy is
increased, this ratio is reduced significantly but does not reach
zero, indicating that some part of the optically oriented holes
retain their spin orientation during energy relaxation. For all values
of the detuning, we also extract the trion (electron) spin decoherence
time $\tau_\mathrm{t}=1/\gamma_{\mathrm{t}}$, and the hole spin dephasing
time, $\tau_\mathrm{h}=1/\gamma_{\mathrm{h}}$, from the fits to
experimental data using Eq.~\eqref{kerr}. Their values, depicted in
Fig.~\ref{TRKR_lambda}(c), remain nearly constant throughout the
investigated detuning range, indicating that the photocarrier and hole
spin dynamics are not strongly influenced by the initial energy
relaxation of non-resonantly excited carriers. It is also clear that
the electron (trion) spin life time remains close to the photocarrier
recombination time. Using TRPL, we measure a photocarrier recombination time of about 175~ps at a temperature of 15~K, using nonresonant  excitation with larger detuning than during the TRKR measurements. As the recombination time increases with detuning~\cite{Bayer_inbook}, this value only provides an upper bound for the photocarrier lifetime under the excitation conditions in the TRKR measurements. We may therefore conclude that the electron spin coherence time is mostly limited by the carrier lifetime for weak, nonresonant excitation.

\begin{figure} \centering
  \includegraphics[width= 0.5\textwidth]{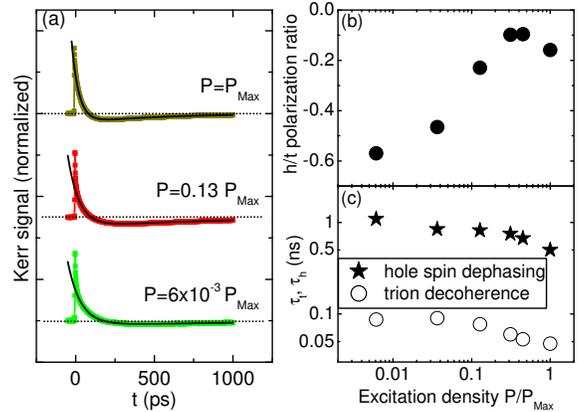}
  \caption{(Color online) (a) TRKR traces measured at 10~K with fixed, near-resonant
    laser excitation energy and various pump powers (symbols). The
    solid lines represent fits to the data according to 
    the theory (extended to $t<0$ for better visibility).  (b) Ratio
    of hole and trion spin polarizations 
    after rapid initial hole dephasing extracted from the measurements
    as a function of excitation density. (c) Hole spin dephasing time
    (solid stars) and trion spin decoherence time (open symbols) as a
    function of excitation density.}
   \label{TRKR_power}
\end{figure}

Next, we discuss power-dependent TRKR measurements. For this series, the
pump power was increased, relative to the values used in the previous
series, by more than two orders of magnitude. Figure \ref{TRKR_power}(a)
shows three TRKR traces for different excitation powers. The laser
excitation energy was chosen to be near-resonant and kept fixed
throughout the series, the sample temperature was 10~K. While for weak
pumping, the TRKR traces show almost no negative part, it becomes
quite pronounced for higher pump powers and the subsequent decay of
the negative signal becomes more rapid. As in the previous measurement
series, our theory closely fits the experimental traces, except for
the first few picoseconds, in which the initial hole spin dephasing
takes place. From the extracted ratio of hole and electron spin
polarization we see that for weak, near-resonant pumping, the hole
spin polarization is significantly larger than for stronger pumping
(Figure \ref{TRKR_power}(b)), most likely indicating the importance of
spin non-conserving
carrier-carrier scattering for rapid hole spin decoherence. We also
observe that the long-time hole and trion spin dephasing times
\emph{decrease} as the pump power is
increased. This decrease is rather weak (by about a factor of 2 over
more than two orders of magnitude of the pulse power) and may be due to
sample heating by the pump beam. A rapid
decrease of the hole spin dephasing time with temperature has been
observed previously by several
groups~\cite{ganichev04,yugova:167402,Korn10}.
The decreasing trion spin lifetime observed in the
power-dependent experiments is not limited by faster photocarrier
recombination, as we observe in TRPL under nonresonant excitation
conditions that the photocarrier recombination time in our sample
\emph{increases} as the  temperature is raised, from 150~ps at 4~K
to 400~ps at 40~K. Such an increase is typically observed in the low
temperature-regime for intrinsic, as well as p- or n-doped QW
structures~\cite{Feldmann87,Ciulin00,olbrich:245329,kugler:035325}.
Therefore, the reduction of the trion spin lifetime must be  caused
by spin-related decoherence processes. Most likely, this is due to
an increased effective \emph{k} vector which leads to larger
spin-orbit fields and more rapid dephasing via the Dyakonov-Perel
mechanism~\cite{DP}.

\subsection{Resonant spin amplification}
\label{sec:results-rsa}

\begin{figure}
\centering
\includegraphics[width=85mm]{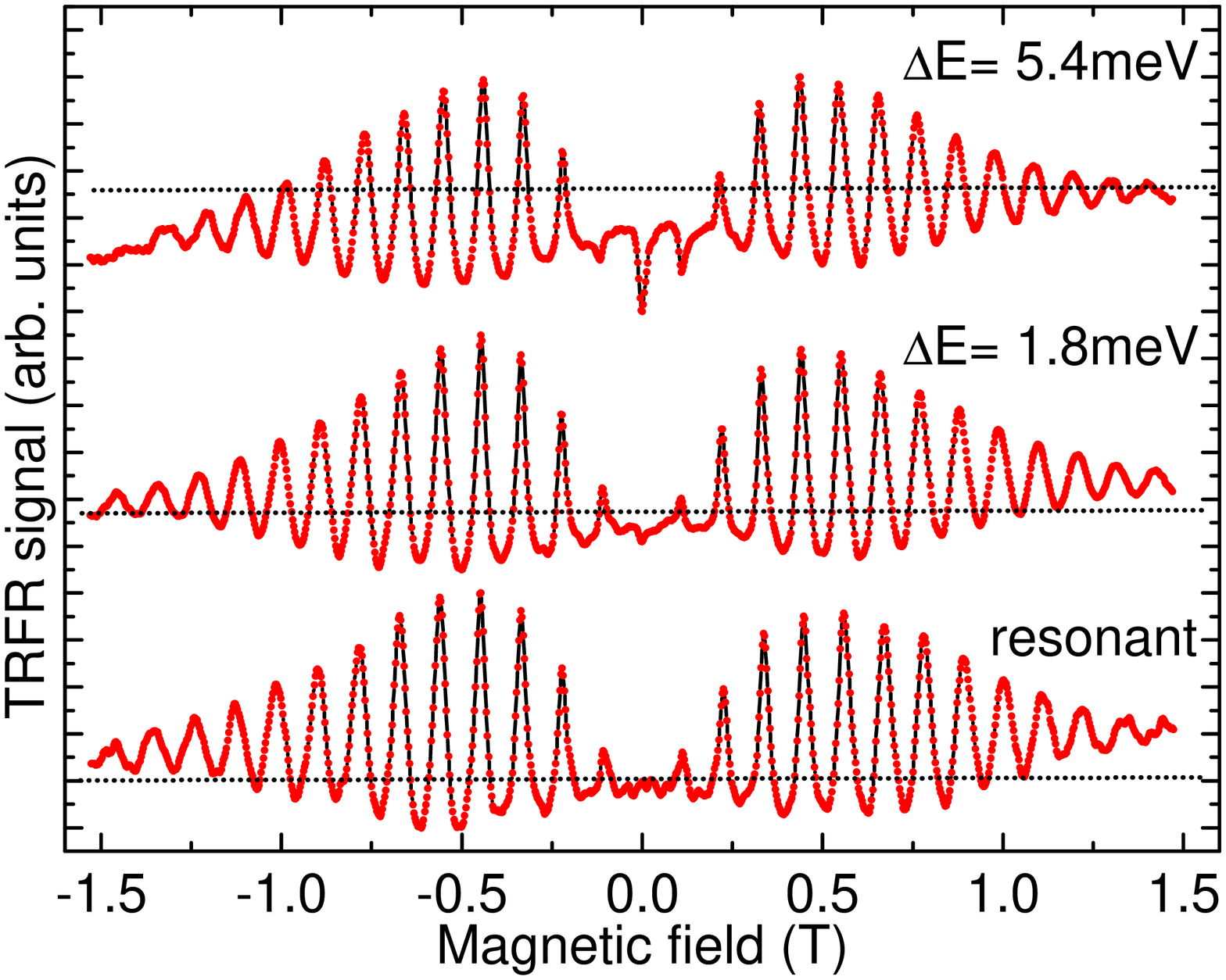}
\caption{\label{fig:exp-wl}(Color online)
The measured RSA traces for resonant and off-resonant excitation
conditions at low excitation power, for three different values of
pump pulse detuning.}
\end{figure}

We now turn to the resonant spin amplification measurements. All
series of measurements were performed at a nominal sample temperature
of 1.2~K in Voigt geometry. In this temperature range, we
previously observed hole spin dephasing times above
70~ns~\cite{Korn10}, which exceed the laser repetition period and lead
to well-defined RSA signals. Figure~\ref{fig:exp-wl}
shows the two principally different shapes of the RSA traces we
observe in experiment:
For resonant excitation conditions (lower curve), in which the
optically oriented hole
spin polarization is conserved during the photocarrier lifetime,
the RSA traces have a
characteristic, batwing-like shape, and the maximum at zero magnetic
field is absent. This peculiar shape arises from the process in which
the spin polarization of optically oriented holes is turned into
the resident hole spin polarization after
recombination~\cite{syperek07,Korn10}: At zero
magnetic field, the optically oriented hole spin polarization is
removed by photocarrier recombination, as described above. With an
applied in-plane magnetic field, however, the strongly different g
factors of electrons and holes lead to incommensurate precession of
the optically oriented spin polarizations, allowing the electrons to
recombine with unpolarized, resident holes, so that some hole spin
polarization remains in the sample after photocarrier
recombination.
This hole spin polarization is oriented in the same way as the
optically created holes (which we will refer to as the
\textit{positive} orientation).
If the hole precession frequency is
an integer multiple of the
laser repetition rate this spin polarization is resonantly
magnified to yield pronounced, equally spaced peaks in the RSA signal.
Their magnitude  grows as the magnetic field increases due to the increased efficiency of the aforementioned initialization process.
The decay of the RSA peak height with further
increase of the magnetic field stems from hole ensemble dephasing due
to the $g$-factor inhomogeneity~\cite{Bayer_inbook}.

For off-resonant excitation, when a substantial part of the optically
oriented hole spin polarization dephases before photocarrier
recombination (as discussed in Sec.~\ref{sec:results-kerr} above), the
RSA signal shape becomes more complex, as the middle and upper traces in
Fig,~\ref{fig:exp-wl} show. Here, we observe RSA
peaks also at zero magnetic field, and the amplitude of these peaks
initially decreases at low fields. For a certain field, we observe a
change of polarity of the RSA peaks, then the RSA peak amplitude first
increases, then decreases again. These observations can be understood
in the following way: For zero magnetic field, a fast partial
re-equilibration of the hole spin polarization leads to negative final
polarization of the resident holes due to removal of the optically
aligned holes by spin-conserving electrons upon recombination, as
discussed in Sec.~\ref{sec:results-kerr}. As the hole spins
are static in the absence of precession at $B=0$, the spin
polarization created in this way by subsequent pump pulses
constructively interferes, leading to the observed negative zero-field
peak.
At non-zero fields, this process competes with the spin alignment due
to the trion precession discussed above, which effectively leads to randomization of the
trion spin orientation at
the moment of recombination so that recombination with an oppositely
oriented hole becomes possible.
As a result of this process, an excess population of
optically oriented holes is left behind after recombination, which
results in a positive spin orientation. Obviously, this mechanism is
only effective at non-zero magnetic fields and its role increases as
the field grows.
This increasing compensation between these two processes is manifested
by reduced negative RSA peak amplitudes at growing
fields, visible in the upper trace in
Fig.~\ref{fig:exp-wl}. At a
certain magnetic field, the precession-related hole spin
initialization process becomes dominant, leading to a positive
spin orientation of the resident holes. Hence, we observe the polarity
change in the RSA
peaks. For even larger magnetic fields, reduced RSA peak height is
again observed due to the inhomogeneous ensemble dephasing.

The competition of the two orientation processes leading to opposite
spin polarizations is reflected in our theory by the factor $f$ in
Eq.~\eqref{RSA} which (apart from the inhomogeneous dephasing)
determines the envelope of the RSA response.
The hole spin relaxation which tends to
re-equilibrate the hole spin polarization after optical creation of
extra holes with positive spin orientation (or,
equivalently, depletion of the negatively oriented holes by binding
them with the optically created excitons into trions) is described by
the first two terms in Eq.~\eqref{RSA1}. Here, according to
Eq.~\eqref{fast-relax}, $e^{-u}$ is the degree of the fast spin
polarization decay during or just after the excitation. The trion
precession, which leads to orientation in the positive
direction, is accounted for by the third term. It is clear from the
form of Eq.~\eqref{RSA1} that a sign change of $f$, corresponding to a
change of the ``polarity'' of the RSA response, is possible only for
$u>0$, that is, in the presence of initial spin relaxation. Moreover,
the position of this transition is shifted to higher fields as $u$
increases, leading to a growing number of inverted RSA peaks in the
low field region.

The growing number of inverted peaks is clearly seen in
Fig.~\ref{fig:exp-wl}: For
increasing laser detuning, first a single negative RSA peak at zero
field develops, then additional peaks of the same orientation are
seen, so that the magnetic field for which a crossover between the
initialization mechanisms occurs is increased. This allows us to
conclude that the increasing detuning of the pump pulse towards higher
energies leads to increased initial spin relaxation. This is in fact
expected, as off-resonant excitation supplies extra energy to the
system leading to additional relaxation processes that usually take
place on picosecond time scales and can lead to spin flips.

\begin{figure}
\centering
\includegraphics[width=85mm]{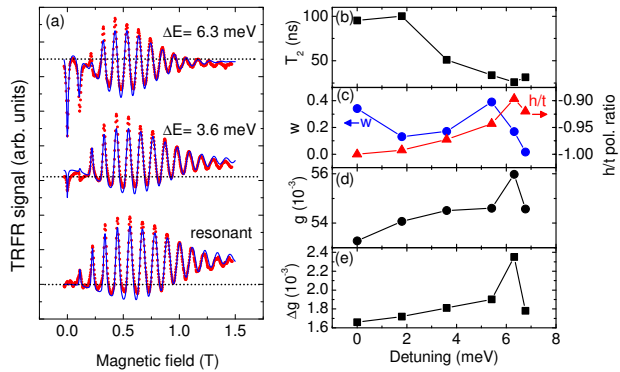}
\caption{\label{fig:rsa-wavelength}(Color online)
(a) Experimental RSA traces (red points) and best fits
according to Eqs.~\eqref{RSA}--\eqref{RSA3} (blue lines) for
selected values of the detuning. (b)--(e) Parameter values extracted
from the fitting: The dephasing time (b), the fast decoherence
parameter and the ratio of hole and electron spin polarization (c), the hole $g$-factor (d), and the standard deviation of
the $g$-factor distribution in the ensemble (e). (the lines are guide to the eye)}
\end{figure}

\begin{figure}
\centering
\includegraphics[width=85mm]{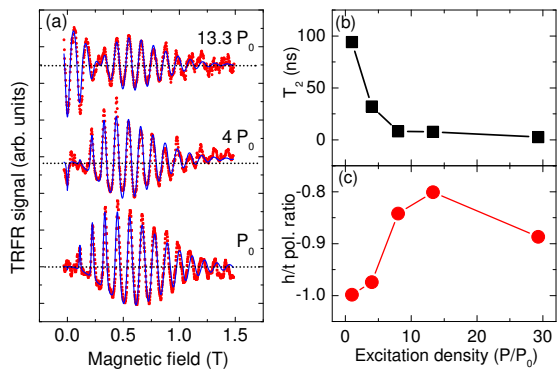}
\caption{\label{fig:exp-pow}(Color online)
(a) Experimental RSA traces (red points) and best fits
according to Eqs.~\eqref{RSA}--\eqref{RSA3} (blue lines) for
selected values of the excitation density. Parameter values extracted
from the fitting as a function of excitation density: (b) Dephasing time $T_2$. (c) Ratio of hole and trion spin polarization.}
\end{figure}

Our theory allows us to closely model the RSA signal shape
using Eq.~\eqref{RSA} integrated over the inhomogeneous distribution
of $g$-factors. In the theoretical modeling underlying the fitting, we
assume that the intrinsic dephasing rates are
constant in the relevant range of the magnetic field. This amounts to
assuming that the spectral densities of the reservoir coupled to the
hole spins are constant in the corresponding range of
frequencies. This is true in particular for an Ohmic reservoir in the
high temperature regime, which is the case here as
$\hbar\wh\ll k_{\mathrm{B}}T$ in the whole range of magnetic fields studied.
Examples of least squares fits obtained in this way for the positive-field
parts of some of the measured RSA traces are
presented in Fig.~\ref{fig:rsa-wavelength}(a). The modeling result (blue
line) not only reproduces all the features of the experimental RSA traces (red
points) but also yields values that are quantitatively
close to the measurement results. This good agreement allows us to
extract the values of various physical characteristics of the hole spin
system. In Fig.~\ref{fig:rsa-wavelength}(b), we show the value of
the transverse spin dephasing time,
$T_{2}=2/\kappa$. This intrinsic dephasing time increases for
decreasing detuning and saturates for low detunings at about
100~ns. The intensities of the fast decoherence and
the ratio of hole and trion spin polarization
are plotted in
Fig.~\ref{fig:rsa-wavelength}(c). The ratio of
  hole and trion spin polarization
reaches -1 as the resonance is approached.
This clearly demonstrates that the rapid initial dephasing
induced by the off-resonant excitation at low temperatures leads only
to a small loss of the optical orientation which disappears completely
at resonance.
On the contrary, the
dephasing factor $w$ remains finite even at the resonance. This is due
to the fact that spin
dephasing is induced by an optical excitation due to selective
coupling of the light field
to one of the spin states (according to the selection rules)
\cite{grodecka09}.
From our fitting we conclude that the $g$-factor tends to increase
slightly with growing detuning (Fig.~\ref{fig:rsa-wavelength}(d)). The
same holds true for the standard
deviation of the ensemble $g$-factor distribution
(Fig.~\ref{fig:rsa-wavelength}(e)). Both of these
  effects may be explained by a reduced absorption of the pump pulse
  as the detuning is increased, leading to a reduction of the sample
  temperature, and a smaller spin-polarized hole ensemble. A shift of
  the hole $g$ factor to larger absolute values with temperature
  reduction has already been observed by Syperek
  et. al.~\cite{syperek07} A smaller ensemble of spin-polarized holes
  is more susceptible to $g$ factor inhomogeneity which arises from
  local fluctuations of, e.g., the QW width or the disorder
  potential.

The theoretical curves turn out to
be sensitive to all the model parameters except for $\kappa'$. The
latter affects the shape of the curves only at very low magnetic
fields since, according to Eqs.~\eqref{RSA2} and \eqref{RSA3}, it can
be neglected when $\kappa'\ll \wh$ which holds already in the vicinity
of the first peak. The values of $\kappa'$ obtained for non-zero
detuning range from 0.012 to 0.03~ns$^{-1}$, which corresponds to
about 50\% of the value of $\kappa=2/T_{2}$ (except for the largest
detuning, where $\kappa'$ is lower). This suggests that the spin
decoherence is dominated by the in-plane dephasing, described by the
rate $\kappa_{z}$, as opposed to the relaxation of the projection on
the structure normal (described by $\kappa_{x}$). This is expected for
heavy holes, as the relaxation of the axial component would involve a spin
transfer of $3\hbar$ and therefore should be suppressed.

As can be
seen in Fig.~\ref{fig:exp-pow}(a),  similar effects in the RSA traces are  observed for
increasing pump power at resonant excitation conditions:  first, a zero-field
peak appears in the RSA traces, then additional peaks are observed for
higher pump powers. Again, our theory allows us to precisely model the
shape of the experimental RSA traces and to extract the fast dephasing
parameters and the hole spin dephasing time. Here, we attribute the
growing fast dephasing, indicated by the reduction of the ratio of
hole and trion spin polarization (Fig.~\ref{fig:exp-pow}(c)), to
the considerably increased amount of energy pumped into the
system. This leads to an increased density of various excitations and,
in consequence, to stronger spin non-conserving scattering, in the
time window before photocarrier recombination takes place, as observed
in the TRKR measurements. An increase of the pump power also
influences the spin dephasing time, $T_2$, of the resident hole spins,
i.e., the spin dynamics after photocarrier recombination, most likely
due to sample heating, as can be seen in the drastic reduction of
$T_2$ (Fig.~\ref{fig:exp-pow}(b)). This interpretation is supported by
the fact that the hole $g$ factor slightly \emph{decreases} with
increasing pump power (not shown), as expected for an increasing
sample temperature.

\section{Conclusion}
\label{sec:concl}

We have performed a time-resolved study of hole spin dynamics in a
p-modulation doped quantum well under different excitation
conditions. In time-resolved Kerr rotation
measurements at zero magnetic field, we observed
the appearance of a hole spin polarization oriented opposite to the
optically oriented holes for non-resonant or high-intensity excitation
conditions. In
resonant spin amplification measurements under non-resonant or
high-intensity excitation conditions, we observe a competition between
two
different initialization processes for a resident hole spin
polarization, which leads to a complex shape of the RSA
traces and to the appearance of inverted spin polarization at low
magnetic fields. Negative spin orientation was earlier observed
in luminescence from $n$-doped quantum dot systems under off-resonant
excitation 
\cite{bracker05,laurent06,shabaev09}.  Our Kerr rotation and RSA
results show  that it is possible not only to 
optically polarize hole spins in a $p$-doped system but also to control the
sign of this polarization by changing \textit{either} the excitation
conditions \textit{or} the magnetic field.

We developed a theoretical model which quantitatively
describes the time-resolved Kerr and RSA signals and allows us to
attribute the spin polarization at zero field to a
decoherence-assisted process in which the hole spin polarization
partly
relaxes towards equilibrium within a very short time after a high-power
or off-resonant excitation. The very good agreement obtained between
the measurement data and the model results allows us to extract
the parameters relevant to the hole spin dynamics, including  the ratio
of hole and electron spin polarizations after optical orientation,
the intrinsic (homogeneous) spin coherence time $T_{2}$, and the
$g$-factor distribution in the ensemble.

Remarkably, rapid initial hole spin dephasing on the
few-picosecond timescale and long hole spin dephasing times reaching
a hundred nanoseconds coexist under off-resonant excitation conditions
in low magnetic fields at low temperatures. Thus, our findings open
the way to optical spin orientation under conditions that assure a
long lifetime of the oriented hole spins.

This work was supported in parts by the DFG (Germany) under SPP 1285 and
SFB 689 (M.~Kugler, M.~Griesbeck, T.~Korn, C.~Sch\"uller), by the
Foundation for Polish Science under the TEAM programme,
co-financed by the European Regional Development Fund,
(K.~Korzekwa and P.~Machnikowski),
and by a Research Group Linkage Project of the Alexander
von
Humboldt Foundation (P.~Machnikowski and
T.~Kuhn). The authors would like to thank
M.~Glazov for highly fruitful discussion.


\end{document}